\documentclass[12pt]{article}
\newcommand{\be}{\begin{equation}}
\newcommand{\ee}{\end{equation}}
\newcommand{\bea}{\begin{eqnarray}}
\newcommand{\eea}{\end{eqnarray}}
\usepackage{graphicx}
\begin{document}
\title{Channel fidelities for high-fidelity approach in KLM scheme
  }
\author{Kazuto Oshima\thanks{E-mail: oshima@nat.gunma-ct.ac.jp}    \\ \\
\sl National  Institute of Technology,  Gunma College,  Maebashi 371-8530, Japan }

\date{}
\maketitle
\begin{abstract}
We study channel fidelity for the high-fidelity approach in the Knill-Laflamme-Milburn (KLM) scheme.
We examine an optimal channel fidelity $f_{opt}$ and identify the corresponding KLM ancilla state. 
In the limit of large $n$, where $2n$ is the number of the ancilla qubits, we find $f_{opt}=1-{\pi^{2} \over 
6n^{2}}+{2\pi^{2} \over 9n^{3}}$.  We see that as $n$ increases $f_{opt}$ approaches to 1 slightly faster than $f=1-{2 \over n^{2}}$ which is the channel fidelity computed by Franson et. al. in the limit of large $n$.  We also compute the channel fidelity for the ancilla state that gives a lower bound of success probability of quantum teleportation.
\end{abstract}

\newpage
\section{Introduction}
Channel fidelity \cite{Horodecki} is one of characters that represent performance of a quantum communication
channel.  Quantum teleportation\cite{Bennett} takes an important role in quantum communication.
Quantum teleportation also produces promising  strategy in quantum computation\cite{Gottesman, Briegel}.
Photon is easily transmitted far away with scarcely being affected with noise in ordinary temperature.
Therefore, photon is one of hopeful media of quantum information.  We can quantum teleport a photon
only probabilistically\cite{Zeilinger}. Knill, Laflamme and Milburn(KLM) \cite{Knill} have invented a scheme to
quantum teleport a photon with success probability near to 1 by introducing an adequate $2n$-qubits ancilla
state that is called a KLM state.  Franson et. al.\cite{Franson} have proposed an approach to improve the success probability in the sense of fidelity by tuning the KLM ancilla state.  Their result, however, depends on the large $n$ analysis, and no particular ancilla state is given concretely. 

The purpose of this paper is to study optimal channel fidelity for the high fidelity approach in the KLM
scheme.  We also give the corresponding ancilla state.   The state $|+\rangle={1 \over \sqrt{2}}(|0\rangle
+|1\rangle)$ is the most difficult state to quantum teleport for the high fidelity approach\cite{Oshima}. 
We identify the ancilla state that maximizes the success probability for $|+\rangle$.  We also evaluate
the channel fidelity for this ancilla state.   We exhibit a simple optical circuit that probabilistically
produces an intended ancilla state from the original KLM state\cite{Imoto}.   Preparing KLM-type ancilla states
has already been discussed in the literature \cite{Franson2, Lemr}.

\section{Channel fidelities} 
We prepare a $2n$-qubits ancilla state as $|t_{n}\rangle=\Sigma_{i=0}^{n}{c(i)}|0\rangle^{n-i}|1\rangle^{i}
|0\rangle^{i}|1\rangle^{n-i}$, where $|0\rangle^{i}$ means $i$ photons in the state $|0\rangle$ etc. and  $c(i)$'s are real coefficients normalized as $\Sigma_{i=0}^{n}c(i)^{2}=1$.  It is convenient to introduce a vector
${}^{t}{\bf c}={}^{t}(c(0),c(1), \cdots, c(n+1))$.
In the original KLM scheme they are settled as $c(i)={1 \over \sqrt{n+1}},i=0,1,\cdots n$. We consider to teleport a quantum state $|\psi\rangle=\alpha|0\rangle +\beta|1\rangle=e^{i\gamma}(\cos{\theta \over 2}|0\rangle+e^{i\varphi}\sin{\theta \over 2}|1\rangle), |\alpha|^{2}+|\beta|^{2}=1$. We perform $n+1$-point quantum Fourier transformation ${\hat F}_{n+1}$ on the state $|\psi\rangle$ and the first $n$ qubits in the ancilla state.  Suppose we observe $k(0 \le k \le n+1)$ photons
after the transformation.  When $k=0$ and $k=n+1$ we loose the original state  $|\psi\rangle=\alpha|0\rangle +\beta|1\rangle$ completely.   In another case, we obtain the quasi teleported state  
\begin{equation}
 |q_{k}\rangle={{\alpha}c(k)|0\rangle+{\beta}c(k-1)|1\rangle \over \sqrt{|\alpha|^{2}c(k)^{2}+|\beta|^{2}c(k-1)^{2}}}
\label{kstate}
\end{equation}
at the $k$-th qubit in the latter half of the ancilla qubits.   To obtain the state $|q_{k}\rangle$ in the form of Eq.(1), we should perform certain relative phase shift between the states $|0\rangle$ and $|1\rangle$ depending
on the observed $k$-photon state.  The probability $p_{k}$
to obtain the state $|q_{k}\rangle$ is given by
\begin{equation}
p_{k}=\Sigma_{k}| \langle k|{\hat F_{n+1}}|\psi\rangle|t_{n}\rangle|^{2}=|\alpha|^{2}c(k)^{2}+|\beta|^{2}c(k-1)^{2},
\label{kprobability}
\end{equation}
where the summation about $k$ runs over all possible $k$-photon states and we have used $\Sigma_{k}|k\rangle\langle{k}|={\hat I}_{k}$ with ${\hat I}_{k}$ the identity operator on any $k$-photon state.  In the
high-fidelity approach the success probability $p_{\bf c}(|\psi\rangle)$ is defined by the expectation value of the square of the fidelity $|\langle \psi|q_{k}\rangle|^{2}, k=1,2,\cdots, n$;  $p_{\bf c}(|\psi\rangle)$ is defined as
$p_{\bf c}(|\psi\rangle)=\sum_{k=1}^{n}p_{k}|\langle \psi|q_{k}\rangle|^{2}$.
Therefore, the success probability for the state  $|\psi\rangle=\alpha|0\rangle+\beta|1\rangle$ is given by
\begin{equation}
p_{\bf c}(|\psi\rangle)=\Sigma_{k=1}^{n}(|\alpha|^{2}c(k)+|\beta|^{2}c(k-1))^{2}.
\label{probability}
\end{equation}

 Channel fidelity $f_{\bf c}$ is defined by
 $f_{\bf c}=\int d\psi p_{\bf c}(|\psi\rangle)={1 \over 4\pi}\int_{0}^{2\pi}d\varphi\int_{0}^{\pi}d{\theta}\sin{\theta}p_{\bf c}(|\psi\rangle)$ that is an average over all input pure states uniformly distributed on the Bloch sphere surface. 
Using $\int_{0}^{\pi}\cos^{4}{\theta \over 2}\sin{\theta}d{\theta}=\int_{0}^{\pi}\sin^{4}{\theta \over 2}\sin{\theta}d{\theta}={2 \over 3}$ and $\int_{0}^{\pi}\cos^{2}{\theta \over 2}\sin^{2}{\theta \over 2}\sin{\theta}d{\theta}
={1 \over 3}$, $f_{\bf c}$ is given by $f_{\bf c}={}^{t}\bf{c}{\tilde A}{\bf c}$, where ${\tilde A}$ is the following
$(n+1) \times (n+1)$ matrix
\begin{eqnarray}
{\tilde A}={1 \over 3}\left(\begin{array}{cccccc}
       1 & {1\over 2} & 0 & \ldots &  0 &0\\
       {1\over 2} & 2 & {1 \over 2} &  \ddots& 0 & 0\\
       0  & {1\over 2} & 2  &  \ddots &  0 & \vdots    \\
       \vdots  & \ldots & \ddots  & \ddots & {1\over 2}&0 \\
       0 & \ldots & \ldots  & {1\over 2} & 2 & {1 \over 2}  \\
       0 &  \ldots &    \dots & 0 & {1\over 2} & 1    \end{array} \right). 
\end{eqnarray}
Let us  $\mu_{n}$ be the largest eigenvalue of ${\tilde A}$ and ${\bf u}$ be the corresponding 
normalized eigenvector.   Setting ${\bf c}$ to be ${\bf u}$, we obtain an optimal fidelity 
$f_{opt}=f_{\bf u}={\mu_{n}}$.

The eigenvalues and the eigenvectors of the matrices ${\tilde A}$  have been studied by
Yueh\cite{Yueh}.
The eigenvector ${}^{t}{\bf u}={}^{t}(u(0),u(1),\cdots,u(n))$ is given by $u(j)=u(0)(\sin(j+1)\theta+2\sin{j}\theta)/\sin{\theta}$, where $\theta(>0)$ is the smallest angle satisfying
\begin{equation}
{1 \over 4}\sin(n+2)\theta+\sin(n+1)\theta+\sin{n}\theta=0.
\label{theta}
\end{equation}
Using $\theta$, the largest eigenvalue $\mu_{n}$can be written as\cite{Yueh} 
\begin{equation}
\mu_{n}={2 \over 3}+{1 \over 3}\cos{\theta}.
\end{equation}
From the normalization condition $u(0)^{2}$ is given by
\begin{equation}
u(0)^{2}={\sin^{2}{\theta} \over \Sigma_{j=0}^{n}(\sin(j+1)\theta+\sin{j}\theta)^{2}}.
\end{equation}

We introduce the following $(n+1) \times (n+1)$ matrix $A$\cite{Oshima} that has some nice properties
\begin{eqnarray}
A={1 \over 4}\left(\begin{array}{cccccc}
       1 & 1 & 0 & \ldots &  0 &0\\
       1 & 2 & 1 &  \ddots& 0 & 0\\
       0  & 1 & 2  &  \ddots &  0 & \vdots    \\
       \vdots  & \ldots & \ddots  & \ddots & 1&0 \\
       0 & \ldots & \ldots  & 1 & 2 & 1  \\
       0 &  \ldots &    \dots & 0 & 1 & 1    \end{array} \right). 
%\label{matrix}
\end{eqnarray}

The largest eigenvalue of $A$ is given by $\lambda_{n}={1 \over 2}+{1 \over 2}\cos{\pi \over n+1}$ and
the  corresponding eigenvectors is denoted as ${}^{t}{\bf v}={}^{t}(v(0),v(1),\cdots,v(n))$, where
$v(j)=v(0)(\sin(j+1){\pi \over n+1}+\sin{j}{\pi \over n+1})/\sin{\pi \over n+1}$. From the normalization
condition $v(0)^{2}$ is given by
\begin{equation}
v(0)^{2}={\sin^{2}{\pi \over n+1} \over \Sigma_{j=0}^{n}(\sin(j+1){\pi \over n+1}+\sin{j}{\pi \over n+1})^{2}}.
\end{equation}

The two matrices ${\tilde A}$ and $A$ are related as
\begin{equation}
{\tilde A}={2 \over 3}A+{1 \over 3}E-{1 \over 6}\Gamma,
\end{equation}
where $E$ is the $(n+1) \times (n+1)$ identity matrix and  $(n+1) \times (n+1)$ matrix $\Gamma$ is defined by
\begin{eqnarray}
\Gamma=\left(\begin{array}{cccc}
       1 & 0  & \ldots  &0\\
       0 & 0  &  \ldots & \vdots\\
       \vdots  & \ldots & \ddots  &0 \\
       0 &  \ldots  & 0  & 1    \end{array} \right).
\end{eqnarray}
 From the inequality $f_{opt}={}^{t}{\bf u}{\tilde A}{\bf u}
\ge {}^{t}{\bf v}{\tilde A}{\bf v}=f_{\bf v}$ we have the following inequality
\begin{equation}
f_{opt} \ge f_{\bf v}= {2 \over 3}+{1 \over 3}\cos{\pi \over n+1}-{1 \over 3}v(0)^{2},
 \label{fv}
\end{equation}
where we have used $v(0)=v(n)$.   From ${}^{t}{\bf u}{A}{\bf u}\le {}^{t}{\bf v}{A}{\bf v}$  we have another inequality $f_{opt}={}^{t}{\bf u}{\tilde A}{\bf u}=
{2 \over 3}{}^{t}{\bf u}{A}{\bf u}+{1 \over 3}-{1 \over 6}(u(0)^{2}+u(n)^{2}) \le {2 \over 3}\lambda_{n}+{1 \over 3}
-{1 \over 6}(u(0)^{2}+u(n)^{2})$ that means
\begin{equation}
f_{opt} \le {2 \over 3}+{1 \over 3}\cos{\pi \over n+1}-{1 \over 3}u(0)^{2},
 \label{fu}
\end{equation}
where we have used $u(0)=u(n)$ which should be hold from the symmetric property of ${\tilde A}$ and from that $\mu_{n}$ is the largest eigenvalue.

\section{Large $n$ analyses}

Since the coefficients of Eq.(\ref{theta}) are not symmetric, the value $\theta$ shifts from ${\pi \over n+1}$. 
We can denote the angle $\theta$ as $\theta={\pi \over n+1}+\delta(n)$, where
$\delta(n)$ is expected to be $O({1 \over n^{2}})$ in the limit of large $n$. Substituting $\theta={\pi \over n+1}+\delta(n)$ into Eq.(\ref{theta}), we find $\delta(n)={\pi \over 3n^{2}}$ in the limit of large $n$.
Therefore, in this limit we have up to $O({1 \over n^{3}})$
\begin{equation}
f_{opt}=1-{\pi^{2} \over 6n^{2}}+{2\pi^{2} \over 9n^{3}}.
\label{fopt}
\end{equation} 
Our result satisfies $f_{opt}>1-{2 \over n^{2}}$.   The right hand side of this inequality is the result by
Franson et. al. \cite{Franson}, where the coefficients $c(i)$'s are not specified explicitly. 
In the limit of large $n$, $v(0)^{2}$ is estimated as
\begin{equation}
v(0)^{2}={({\pi \over n})^{2} \over {4n \over \pi}\int_{0}^{\pi}\sin^{2}xdx}={\pi^{2} \over 2n^{3}}.
\end{equation}
In the same way we find
\begin{equation}
u(0)^{2}={({\pi \over n})^{2} \over {9n \over \pi}\int_{0}^{\pi}\sin^{2}xdx}={2\pi^{2} \over 9n^{3}}.
\label{u1}
\end{equation}
These results Eqs.(\ref{fopt})-(\ref{u1}) accord with the inequalities Eqs.(\ref{fv}), (\ref{fu}).
We have $f_{opt}=f_{{\bf v}}+{\pi^{2} \over 18n^{3}}$ in the limit of  large $n$.

We consider to teleport the state $|+ \rangle={1 \over \sqrt{2}}(|0\rangle+|1\rangle)$ that is the most
difficult state to teleport\cite{Oshima}.  Setting the coefficient vector ${\bf c}$ to be the optimal one
${\bf u}$, we have the following success probability 
\begin{equation}
p_{{\bf u}}(|+\rangle)={ 1\over 2}\Sigma_{i=0}^{n}(u(i)+u(i-1))^{2} 
                               ={}^{t}{\bf u }A{\bf u}
                               ={}^{t}{\bf u }({3 \over 2}{\tilde A}-{1 \over 2}E+{1 \over 4}\Gamma) {\bf u}.
\end{equation}
In the limit of large $n$, $p_{{\bf u}}(|+\rangle)$ is estimated as
\begin{equation}
p_{{\bf u}}(|+\rangle)=\lambda_{n}-{\pi^{2} \over 12n^{3}}.
\end{equation}
Therefore, we have, at least at large $n$, $p_{{\bf u}}(|+\rangle) < p_{{\bf v}}(|+\rangle)=\lambda_{n}$
as it should be, because, ${\bf v}$ is the vector that maximizes $p_{{\bf c}}(|+\rangle)$ \cite{Oshima}.

\section{State preparation}
Here we show a simple way how to  prepare the state $\Sigma_{j=0}^{n}c(j)|1\rangle^{j}|0\rangle^{n-j}|0\rangle^{j}|1\rangle^{n-j}$ starting from the original KLM state 
${1 \over \sqrt{n+1}}\Sigma_{j=0}^{n}|1\rangle^{j}|0\rangle^{n-j}|0\rangle^{j}|1\rangle^{n-j}$
by using $2[{n \over 2}]$ beam splitters \cite{Imoto}
 as in Fig.1.
The transmission coefficient $t_{i}$ is settled as $t_{i}={c([{n \over 2}]-i) \over c([{n \over 2}]+1-i)},
i=1,2,\cdots,[{n \over 2}]$. If the $2[{n \over 2}]$ detectors detect no photons
we obtain the state $\Sigma_{j=0}^{n}c(j)|1\rangle^{j}|0\rangle^{n-j}|0\rangle^{j}|1\rangle^{n-j}$ as output.
The success probability is given by ${1 \over (n+1)(c([{n \over 2}]))^{2}}$, which is at least larger
than ${1 \over n+1}$. When we use ${\bf v}$ as ${\bf c}$,
$c([{n \over 2}])^{2}$ is estimated as ${2 \over n}$ in the limit of  large $n$ .  This will mean that we can obtain the intended state with the probability around ${1 \over 2}$.  
\\
\begin{figure}[htbp]
\includegraphics[width=0.5\linewidth]{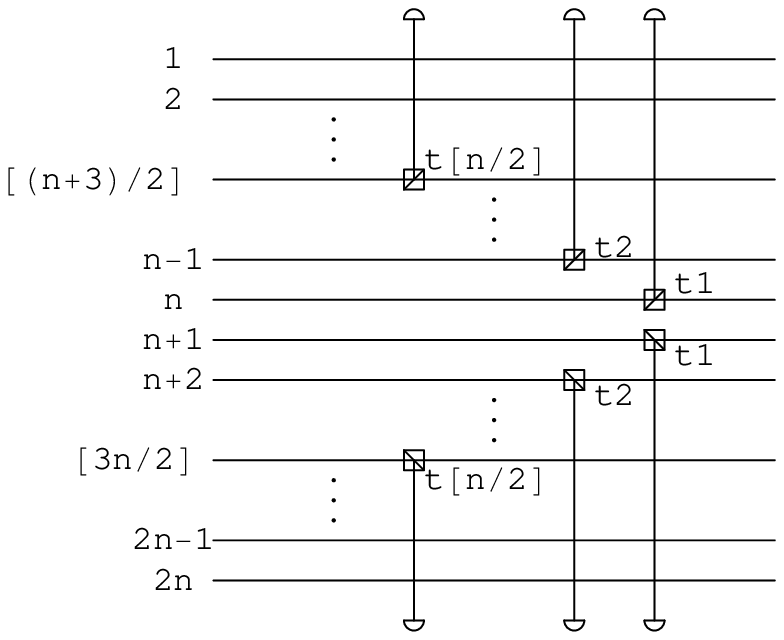}
\\
Fig.1  A circuit of linear optics consisting of $2[{n \over 2}]$ beam splitters and corresponding photo detectors that probabilistically produces the intended  state from the original KLM state.
\end{figure}

\section{Conclusions}
We have examined the optimal fidelity $f_{opt}={2 \over 3}+{1 \over 3}\cos{\theta}$, where $\theta$
is the smallest angle satisfying Eq.(\ref{theta}), for the high fidelity approach in the KLM scheme.  
We have identified the corresponding ancilla state.  In the limit of
large $n$ we have found $f_{opt}=1-{\pi^{2} \over 6n^{2}}+{2\pi^{2} \over 9n^{3}}$ which slightly
exceeds the result $f=1-{2 \over n^{2}}$ by Franson et. al..  We have examined another ancilla state
that gives the maximal success probability for the state $|+\rangle$, which is the most difficult state 
to quantum teleport in the high fidelity approach. For this ancilla state  we have the channel fidelity
$f_{\bf v}={2 \over 3}+{1 \over 3}\cos{\pi \over n+1}-{1 \over 3}v(0)^{2}$, where $v(0)$ is the first
coefficient of the ancilla state.  In the limit of  large $n$ we have $f_{opt}=f_{\bf v}+{\pi^{2} \over 18n^{3}}$.  We also have exhibited an optical circuit producing a required ancilla state starting from the original KLM state.
\\
\\
{\bf Acknowledgment}\\
The author thanks Prof. S.Strelchuk for making him acquainted with the reference \cite{Yueh}.


\newpage
\begin{thebibliography}{99}

\bibitem{Horodecki}
M.Horodecki, P.Horodecki and R.Horodechi, Phys.Rev.A, {\bf 60}, 1888(1999).

\bibitem{Bennett}
        C.H.Bennett, G.Brassard, C.Cr${\acute{\rm  e}}$peau, R.Jozsa, A.Peres and W.K.Wootters, Phys.Rev.Lett.{\bf 70}, 1895(1993).

\bibitem{Gottesman}
D.Gottesman and I.L.Chuang,  Nature, {\bf 402}, 390(1999).
 
\bibitem{Briegel}
 R.Raussendorf and H.J.Briegel, Phys.Rev.Lett., {\bf 86}, 5188(2001);
 H.J.Briegel and  R.Raussendorf, Phys.Rev.Lett., {\bf 86},  910(2001).  

\bibitem{Zeilinger}
D.Bouwmeester, J. Pan, K. Mattle, M. Eibl, H. Weinfurter, A. Zeilinger, Nature{\bf 390},575 (1997).  

\bibitem{Knill}
 E.Knill , R.Laflamme and G.J.Milburn, Nature, {\bf 409}, 46(2001).

%\bibitem{Grudka}
%A.Grudka and J.Modlawska, Phys.Rev.A{\bf 77}, 01431(2008).

\bibitem{Franson}
J.D.Franson, M.M.Donegan, M.J.Fitch, B.C.Jacobs and T.B.Pittman, Phys.Rev.Lett., {\bf 89}, 137901(2002).

\bibitem{Oshima}
K.Oshima,  arXiv: quant-ph/1712.01119.

\bibitem{Imoto}
T.Yamamoto, K.Tamaki, M.Koashi and N.Imoto, Phys.Rev.A{\bf 66}, 064301(2002).

\bibitem{Franson2}
J.D.Franson, M.M.Donegan and B.C.Jacobs, Phys.Rev.A{\bf 69}052328(2003).

\bibitem{Lemr}
K.Lemr, J.Phys.B{\bf 44},195501(2011);\\
K.Lemr, K.Bartkiewicz and A.Cernoch, J.Opt.{\bf 17},125202(2015).

\bibitem{Yueh}
W-C.Yueh, Applied Mathematics E-notes, {\bf 5}, 66(2005).


\end{thebibliography}
\end{document}